%% file: main.tex
\documentclass{article}
\usepackage{arxiv}

\usepackage[utf8]{inputenc} 
\usepackage[T1]{fontenc}    
\usepackage{hyperref}       
\usepackage{url}            
\usepackage{booktabs}       
\usepackage{microtype}      
\usepackage{lipsum}	    	
\usepackage{graphicx}
\usepackage{doi}
\usepackage[section]{placeins}
\usepackage{mathtools}
\usepackage{amsmath}
\usepackage{amssymb}
\usepackage[ruled,vlined]{algorithm2e}
\usepackage{lineno}
\usepackage[dvipsnames]{xcolor}
\usepackage{enumitem}
\usepackage[sort&compress,numbers]{natbib}
\setlist[itemize]{noitemsep, topsep=0pt}
\usepackage{color}
\usepackage{soul}

\title{\textbf{DeepOKAN: Deep Operator Network Based on Kolmogorov Arnold Networks for Mechanics Problems}}

\author{{\hspace{1mm}Diab W. Abueidda}\thanks{da3205@nyu.edu} \\
	Civil and Urban Engineering Department\\
	New York University Abu Dhabi\\
        National Center for Supercomputing Applications\\
	University of Illinois at Urbana-Champaign\\
	\And
        {\hspace{1mm}Panos Pantidis}\\
	Civil and Urban Engineering Department\\
	New York University Abu Dhabi\\
        \And
	{\hspace{1mm}Mostafa E. Mobasher}\thanks{mostafa.mobasher@nyu.edu}\\
	Civil and Urban Engineering Department\\
	New York University Abu Dhabi\\
}

\renewcommand{\shorttitle}{DeepOKAN}

\begin{document}

\maketitle

\begin{abstract}
The modern digital engineering design often requires costly repeated simulations for different scenarios. The prediction capability of neural networks (NNs) makes them suitable surrogates for providing design insights. However, only a few NNs can efficiently handle complex engineering scenario predictions. We introduce a new version of the neural operators called DeepOKAN, which utilizes Kolmogorov Arnold networks (KANs) rather than the conventional neural network architectures. Our DeepOKAN uses Gaussian radial basis functions (RBFs) rather than the B-splines. RBFs offer good approximation properties and are typically computationally fast. The KAN architecture, combined with RBFs, allows DeepOKANs to represent better intricate relationships between input parameters and output fields, resulting in more accurate predictions across various mechanics problems. Specifically, we evaluate DeepOKAN's performance on several mechanics problems, including 1D sinusoidal waves, 2D orthotropic elasticity, and transient Poisson's problem, consistently achieving lower training losses and more accurate predictions compared to traditional DeepONets. This approach should pave the way for further improving the performance of neural operators.
\end{abstract}


\input{sections/1Intro.tex}
\input{sections/2Methods.tex}
\input{sections/3Results.tex}

\input{sections/4Conclusions.tex}

\input{sections/Acknowledgment.tex}
\input{sections/DataAvail.tex}

\bibliography{mybibfile}

\end{document}

%% file: sections/1Intro.tex
\section{Introduction} \label{intro}

Contemporary science and engineering increasingly depend on advanced physics-based computational models. Numerical simulations offer critical insights for understanding and predicting complex physical phenomena and engineering systems. Real-world problems often involve varying loads, boundary and initial conditions, material properties, and domain geometries. Due to their inherent complexity, multi-physics nature, dimensionality, time dependency, and fidelity requirements, finite element analysis (FEA) models can be computationally intensive, even with high-performance computing platforms \cite{ammosov2023online, efendiev2009multiscale, mobasher2021dual}. Consequently, relying solely on traditional high-fidelity simulation models for tasks such as computer-aided design, material discovery, and digital twins is often impractical, especially when exploring numerous design scenarios or geometries \cite{torzoni2024digital}. In this context, surrogate neural network (NN) models emerge as a promising machine learning approach, capable of rapidly inferring solutions to physical problems without requiring expensive numerical simulations once they are trained \cite{kushwaha2023designing, valizadeh2022convolutional, pantidis2024116940}. These models hold significant potential across various application domains, including real-time simulations for predictions and controls, design, topology and shape optimizations, sensitivity analysis, and uncertainty quantification, which require extensive forward evaluations with varying parameters \cite{paermentier2021machine, olivier2021bayesian, parrott2023multidisciplinary}. For a comprehensive overview of the application of neural networks in computational mechanics, see \cite{herrmann2024deep}. In this study, we introduce the DeepOKAN. This novel neural network architecture combines the strengths of Kolmogorov-Arnold networks (KANs) and deep operator networks (DeepONets) to solve complex mechanics problems efficiently. By replacing the traditional multi-layer perceptrons (MLPs) in DeepONets with KANs, we demonstrate improved accuracy compared to existing approaches.

Despite their advantages, most existing surrogate models still necessitate retraining or transfer learning when input parameters such as loads, boundary conditions, material properties, or geometry are altered. This limitation highlights the need for more robust solutions that adapt to changing conditions without extensive reconfiguration. Recently, researchers devised NNs trained to approximate the underlying physics or mathematical operator for a class of problems, a process known as operator learning \cite{li2021neural}. Researchers have proposed various architectures for operator learning, with two notable examples being the Fourier neural operator (FNO) and the deep operator network (DeepONet). The Fourier Neural Operator (FNO) was first introduced by Li et al. \cite{li2020fourier} to solve partial differential equations with parametric inputs. Drawing inspiration from the Fourier transform used in differential equation solutions, the input function is processed through multiple Fourier layers. The encoded information is then mapped onto the output function space. Each Fourier layer applies the fast Fourier transform (FFT) to its input and filters out high-frequency modes. FNO and its enhanced versions have been successfully used for Burger’s equation, Darcy flow, and the Navier-Stokes equation \cite{li2020fourier}, as well as for elasticity and plasticity problems in solid mechanics \cite{li2023fourier}. Nonetheless, since the FNO employs FFT, it encounters challenges when dealing with complex geometries or intricate non-periodic boundary conditions. Additionally, Fourier-based operations are computationally expensive when addressing high-dimensional problems.

Lately, Lu et al. \cite{lu2021learning, lu2021deepxde} introduced the DeepONet, which emerged as another capable architecture for operator learning. DeepONet comprises two sub-networks: a branch network for encoding input functions and a trunk network for encoding input domain geometry. Initially, both networks were designed as MLP networks. In the seminal study, DeepONet effectively mapped between unknown parametric functions and solution spaces for several linear and nonlinear partial differential equations (PDEs) while also learning explicit operators such as integrals. DeepONets have been increasingly used to tackle scientific and engineering challenges, such as the inverse design of nanoscale heat transfer systems \cite{lu2022multifidelity}, brittle fracture \cite{goswami2022physics}, digital twins \cite{kobayashi2024improved}, and uncertainty quantification \cite{garg2023vb}. Lu et al. \cite{lu2022comprehensive} thoroughly compared FNO and DeepONet. Several researchers have proposed different architectures for the branch and/or trunk networks to enhance DeepONet's versatility and capability to capture complex scenarios \cite{he2023novel, he2024sequential, he2024predictions, li2020neural, he2024geom, zhong2024physics}.

The significance of MLPs is immense, as they are the standard models in machine learning for approximating nonlinear functions, owing to their expressive power as assured by the universal approximation theorem \cite{cybenko1989approximation, hornik1989multilayer}. Recently, Liu et al. \cite{liu2024kan} proposed a promising alternative to MLPs called Kolmogorov-Arnold networks (KANs). While MLPs are based on the universal approximation theorem \cite{hornik1989multilayer}, KANs are based on the Kolmogorov-Arnold representation theorem \cite{kolmogorov1961representation, arnol1959representation, braun2009constructive}. Similar to MLPs, KANs have fully connected architectures. Nevertheless, unlike MLPs, which use fixed activation functions on nodes (neurons), KANs employ activation functions with learnable weights on edges. This difference can make KANs more accurate and interpretable than MLPs for several modeling scenarios. The potential of using the Kolmogorov-Arnold representation theorem to construct neural networks has been explored in various studies \cite{sprecher2002space, koppen2002training, lin1993realization, lai2021kolmogorov, fakhoury2022exsplinet} before the work of Liu et al. \cite{liu2024kan}. However, most research has adhered to the original depth-$2$, width-$(2n + 1)$ representation and has not utilized modern techniques such as backpropagation for training. Liu et al. \cite{liu2024kan} generalized the original Kolmogorov-Arnold representation to arbitrary widths and depths, thereby revitalizing and contextualizing it within the contemporary deep learning framework. They also conducted extensive empirical experiments to demonstrate its potential as a foundational artificial intelligence and science model, highlighting its accuracy and interoperability. Although the work of Liu et al. \cite{liu2024kan} was published recently, it has already inspired other researchers to investigate the topic further \cite{genet2024tkan, li2024kolmogorov, ss2024chebyshev}. 

This study introduces the DeepOKAN, where the branch and trunk of DeepONet are KANs rather than the traditional MLPs. In addition to this novel combination, we have integrated Radial Basis Functions (RBFs) for function approximation within the architecture. RBFs have emerged as a powerful tool in numerical methods, providing a flexible and effective way to approximate multivariate functions, especially in the absence of structured grid data \cite{buhmann2000radial, arora2023review}. RBFs are typically used to build up function approximations as a sum of weighted radial basis functions, each associated with a different center point. These functions have been employed in interpolation, approximation, and solving differential equations \cite{arora2023review, chenoweth2012local}. RBFs are advantageous because they provide good approximation properties, handle scattered data, do not require triangulations of the data points offering a meshless approach, and leverage known fundamental solutions (Green's functions) of the underlying PDE operators \cite{buhmann2000radial}. These characteristics make RBFs particularly advantageous in applications involving complex geometries or higher-dimensional problems, where traditional mesh-based methods may be computationally prohibitive. In the context of FEA, RBFs produce shape functions with simplicity and have shown promise in analyzing solid mechanics problems \cite{kien2023radial}. Their application extends to dynamic analysis, where RBF-based approaches have shown promise in capturing transient behavior \cite{elsheikh2023efficient, rashed2002transient}. Also, RBFs have been successfully applied in fracture mechanics, demonstrating their ability to model complex stress fields and crack propagation \cite{wang2010subdomain, wen2007meshless}.

RBFs have been used before in the context of machine learning \cite{satapathy2019empirical}, but their success was limited. However, the recent development of KANs promises more effective integration. By leveraging the theoretical foundations laid by Kolmogorov and Arnold, KANs offer a compact and efficient representation of functions. This efficiency, combined with higher accuracies, positions KANs as a promising alternative to traditional MLPs across various applications. This paper uses the DeepOKAN to solve several mechanics problems, including 1D sinusoidal waves, 2D orthotropic elasticity, and the 2D transient Poisson's problem. The paper's structure is outlined as follows: Section \ref{Methods} provides an overview of KANs and Gaussian RBFs. It also scrutinizes the DeepOKAN. In Section \ref{results}, the DeepOKAN is developed as a neural operator for a few numerical examples. The paper concludes in Section \ref{conclu} by summarizing the key results and indicating potential directions for future research.

%% file: sections/2Methods.tex
\section{Methods}\label{Methods}

\subsection{Kolmogorov-Arnold representation theorem}\label{KAR}

The Kolmogorov-Arnold representation theorem, also known as the superposition theorem, is a fundamental result in approximation theory. It asserts that any continuous multivariate function on a bounded domain can be represented as a composition of a finite number of univariate functions and a set of linear operations (additions). Specifically, for any smooth function $f: [0,1]^{n_{\text{in}}} \rightarrow \mathbb{R}^{n_{\text{out}}}$, the Kolmogorov-Arnold theorem asserts that there exist continuous univariate functions $\psi_{q}$ and $\phi_{p,q}$ such that:

\begin{equation}\label{kar_eq}
\begin{aligned}
   f(\boldsymbol{x}) &= \sum_{q=1}^{2n_{\text{in}}+1} \psi_{q} \left( \sum_{p=1}^{n_{\text{in}}} \phi_{p,q}(x_p) \right),
\end{aligned}
\end{equation}
where $\boldsymbol{x}=(x_{1}, \ldots, x_{d_{\text{in}}})$, $\phi_{q,p} : [0,1] \rightarrow \mathbb{R}$, and $\psi_q : \mathbb{R} \rightarrow \mathbb{R}$.

\subsection{Kolmogorov-Arnold network}\label{KAN}

The Kolmogorov-Arnold representation theorem states that any continuous multivariable function on a bounded domain can be represented as a superposition of continuous functions of one variable and addition. While no strict restrictions exist on choosing these univariate functions beyond their continuity, practical considerations may influence their specific forms based on the problem context and desired properties. In the KAN implementation by Liu et al. \cite{liu2024kan}, each KAN layer comprises the sum of the spline function and the sigmoid linear unit, known as the SiLU activation function, with learnable coefficients. Splines face a significant curse of dimensionality (COD) due to their inability to leverage compositional structures. On the other hand, MLPs are less affected by COD because of their feature learning capabilities. Still, they are less accurate than splines in low-dimensional settings due to their inefficiency in optimizing univariate functions. A model must capture the compositional structure (external degrees of freedom) and effectively approximate univariate functions (internal degrees of freedom) to learn a function accurately. A more detailed description of KANs can be found in the work of \cite{liu2024kan}, and for the sake of completeness, we provide a brief description below.

\begin{figure}[!htb]
    \centering
    \includegraphics[width=0.6\textwidth]{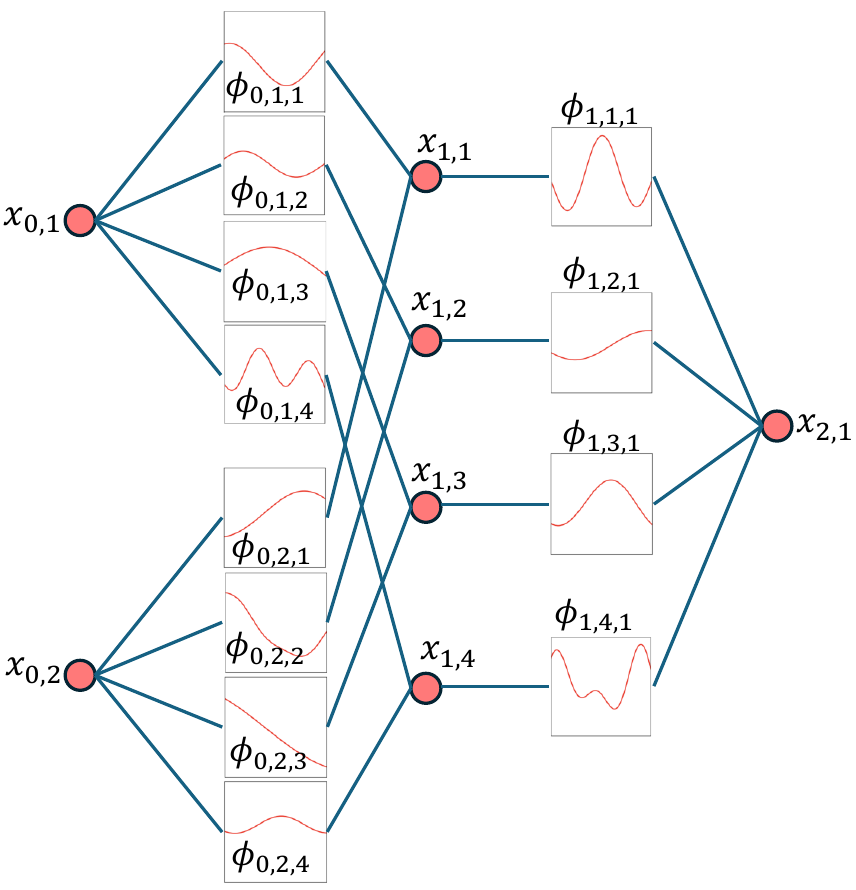}
    \caption{Illustration of the activation functions flowing through the network.}
    \label{KAN_fig}
\end{figure}

A KAN comprises several layers, and in Fig. \ref{KAN_fig}, we show a KAN with two such layers. A KAN layer with $n_{in}$-dimensional inputs and $n_{out}$-dimensional outputs can be represented as a matrix of 1D functions, denoted as:
\begin{equation}\label{psi_mat}
\begin{aligned}
    \boldsymbol{\Psi} = \{ \phi_{l,q,p} \}, \quad p = 1, 2, \ldots, n_{\text{in}}, \quad q = 1, 2, \ldots, n_{\text{out}},
\end{aligned}
\end{equation}
where the functions $\phi_{q,p}$ have learnable parameters. The shape of KAN can be expressed as an integer array $\left[n_{0}, n_{1}, \ldots, n_{L}\right]$, where $n_{l}$ is the number of nodes (neurons) in the $l^\text{th}$ layer (see Fig. \ref{KAN_fig}). The $i^\text{th}$ neuron in the $l^\text{th}$ layer is denoted by $(l,i)$, while the activation value of the $(l,i)$-neuron is represented by $x_{l,i}$. There are $n_{l} n_{l+1}$ activation functions between the consecutive layers $l$ and $l+1$. The activation function connecting the neurons $(l,i)$ and $(l+1, j)$ is denoted by: 
\begin{equation}\label{activation}
\begin{aligned}
    \phi_{l,i,j}, \quad l = 0, \ldots, L-1, \quad i = 1, \ldots, n_l, \quad j = 1, \ldots, n_{l+1}.
\end{aligned}
\end{equation}
As shown in Fig. \ref{KAN_fig}, the activation functions appear on the edges rather than the nodes, unlike MLPs. Specifically, the pre-activation of $\phi_{l,i,j}$ is $x_{l,i}$, while the post-activation of $\phi_{l,i,j}$ is $\Tilde{x}_{l,i,j}$. Then, the $\Tilde{x}_{l,i,j}$'s are summed to determine $x_{l+1,j}$, which is the activation value at the $(l+1,j)$-neuron:
\begin{equation}\label{kan_layer_eq}
\begin{aligned}
    x_{l+1,j} = \sum_{i=1}^{n_l} \Tilde{x}_{l,i,j} = \sum_{i=1}^{n_l} \phi_{l,i,j}(x_{l,i}).
\end{aligned}
\end{equation}

This can be presented in a matrix form as follows:
\begin{equation}\label{kan_layer_mat}
\begin{aligned}
    \boldsymbol{x}_{l+1} = \underbrace{
\begin{pmatrix}
\phi_{l,1,1}(\cdot) & \phi_{l,1,2}(\cdot) & \cdots & \phi_{l,1,n_l}(\cdot) \\
\phi_{l,2,1}(\cdot) & \phi_{l,2,2}(\cdot) & \cdots & \phi_{l,2,n_l}(\cdot) \\
\vdots & \vdots & \ddots & \vdots \\
\phi_{l,n_{l+1},1}(\cdot) & \phi_{l,n_{l+1},2}(\cdot) & \cdots & \phi_{l,n_{l+1},n_l}(\cdot)
\end{pmatrix}}_{\boldsymbol{\Psi}_l} \boldsymbol{x}_l,
\end{aligned}
\end{equation}
where $\boldsymbol{\Psi}_l$ is the function matrix corresponding to the $l^{\text{th}}$ KAN layer. Then, a general KAN is expressed as a composition of $L$ layers:
\begin{equation}\label{KAN_EQ}
\begin{aligned}
\text{KAN}(\boldsymbol{x}) = (\boldsymbol{\Psi}_{L-1} \circ \boldsymbol{\Psi}_{L-2} \circ \cdots \circ \boldsymbol{\Psi}_1 \circ \boldsymbol{\Psi}_0) \boldsymbol{x}.
\end{aligned}
\end{equation}
The original Kolmogorov–Arnold representation (see Equation \ref{kar_eq}) is deemed a special case of the KAB, with two layers and a shape $\left[n_0, 2n_{0} +1, 1\right]$. On the other hand, the well-known MLPs alternate between linear transformation $\boldsymbol{W}$ and nonlinear activation functions $\sigma$: 
\begin{equation}\label{MLP_EQ}
\begin{aligned}
\text{MLP}(\boldsymbol{x}) = (\boldsymbol{W}_{L-1} \circ \sigma \circ \boldsymbol{W}_{L-2} \circ \sigma \circ \cdots \circ \boldsymbol{W}_1 \circ \sigma \circ \boldsymbol{W}_0) \boldsymbol{x}.
\end{aligned}
\end{equation}
MLPs distinctly handle linear transformations $\boldsymbol{W}$ and nonlinearities $\sigma$, whereas KANs combine both within $\boldsymbol{\Psi}$.

\subsection{RBF-KAN}\label{RBFKAN}

\begin{figure}[!htb]
    \centering
    \includegraphics[width=0.7\textwidth]{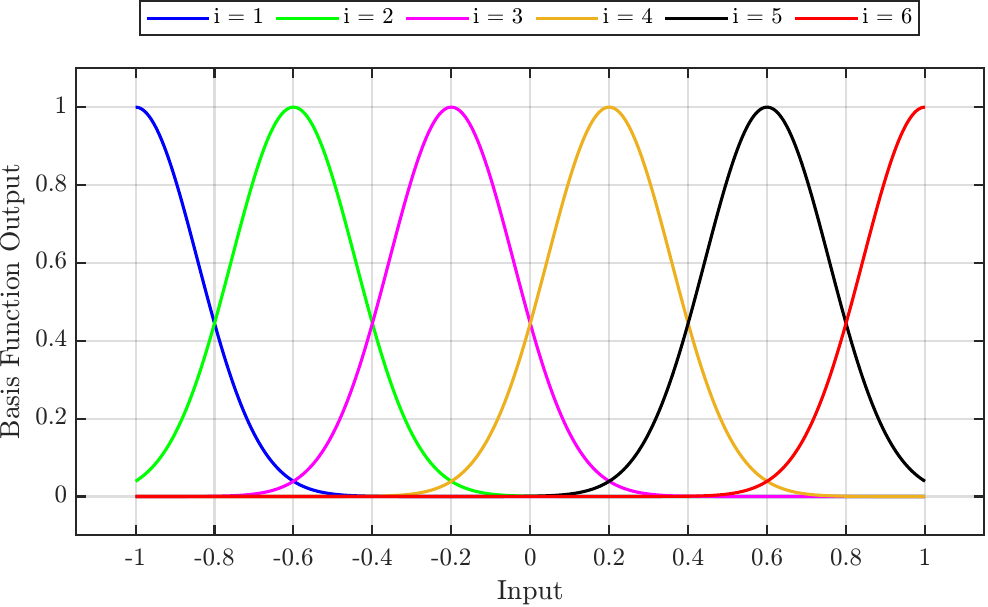}
    \caption{Visualization of RBF-KAN Layer: Each curve represents an individual basis function.}
    \label{Figure_RBFKAN}
\end{figure}

As mentioned earlier, Liu et al. \cite{liu2024kan} proposed using B-splines as activation functions within the KAN framework. Here, we propose using RBFs for KANs. An RBF is a real-valued function whose value depends only on the distance from a central point. The Gaussian RBF is a specific type from the radial basis function family. The transformation of each input $x_{i}$ using a Gaussian RBF $\boldsymbol{R}$ is given by:
\begin{equation}\label{RBF_eq}
\begin{aligned}
R^{l}(x_{i}^{l}, g_{j}^{l}) &= \exp\left(-\left(\frac{x_{i}^{l} - \text{g}_{j}^{l}}{\beta}\right)^2\right), \quad i = 1, 2, \ldots, n, \quad j = 1, 2, \ldots, m
\end{aligned}
\end{equation}
where $n$ is the input dimension (layer $l$ size), and $m$ represents the number of grid points (RBF centers). The superscript $l$ denotes the layer number: $l=1,\ldots, L$, where $L$ is the number of layers.. $g_{j}$ represents the $j^{\text{th}}$ point in the RBF grid; $g_j \in \boldsymbol{G}$. The RBF centers $\boldsymbol{G}$ can be learnable parameters or fixed points (non-learnable) defined during the initialization of the RBF transformation within the RBF-KAN framework. $\beta$ is the scaling factor for the RBF, defined as:
\begin{equation}\label{RBF_beta}
\begin{aligned}
\beta &= \frac{g_{\text{max}} - g_{\text{min}}}{m - 1}. 
\end{aligned}
\end{equation}
$g_{\text{max}}$ and $g_{\text{min}}$ denote the maximum and minimum values of the grid, respectively. After transforming the inputs, the transformed features are combined linearly using some learnable weight matrix $\boldsymbol{W}$. The output $\boldsymbol{x}^{l+1}$ is computed as:
\begin{equation}\label{RBF_out}
\begin{aligned}
\boldsymbol{x}^{l+1} &= \boldsymbol{W}^{l} \boldsymbol{R}^{l}\left(\boldsymbol{x}^{l}, \boldsymbol{G}^{l} \right)
\end{aligned}
\end{equation}
where $\boldsymbol{R}$ is a vector of size $(n\times m,)$, $\boldsymbol{W}$ is a matrix with a size of $(o, n\times m)$, where $o$ is the length of the vector $\boldsymbol{x}^{l+1}$. Fig. \ref{Figure_RBFKAN} illustrates the behavior of the RBF-KAN layer, showcasing a series of Gaussian-like curves, each representing a basis function centered at different points along the input range. These basis functions highlight the localized response characteristic of the RBF layer, which can effectively capture and represent complex, non-linear patterns in the data. Similar curves can be produced using B-spline layers, as shown in the work of Liu et al. \cite{liu2024kan}.

While constructing an RBF-KAN, one can stack several RBF-KAN layers. During the training of an RBF-KAN, $\boldsymbol{W}$ and $\boldsymbol{G}$ are obtained by minimizing a loss function $\mathcal{L}$, which requires finding the gradients of $\mathcal{L}$ with respect to $\boldsymbol{W}$ and $\boldsymbol{G}$ using backpropagation:
\begin{equation}\label{backprop1}
\begin{aligned}
\frac{\partial \mathcal{L}}{\partial \boldsymbol{W}^{l}} &= \frac{\partial \mathcal{L}}{\partial \boldsymbol{x}^{L}} \cdot \boldsymbol{W}^{L-1} \cdot \frac{\partial \boldsymbol{R}^{L-1}}{\partial \boldsymbol{x}^{L-1}} \cdot \boldsymbol{W}^{L-2} \cdot \cdots \cdot \boldsymbol{W}^{l+1} \cdot \frac{\partial \boldsymbol{R}^{l+1}}{\partial \boldsymbol{x}^{l+1}} \cdot \boldsymbol{R}^{l}\\
\frac{\partial \mathcal{L}}{\partial \boldsymbol{G}^{l}} &= \frac{\partial \mathcal{L}}{\partial \boldsymbol{x}^{L}} \cdot \boldsymbol{W}^{L-1} \cdot \frac{\partial \boldsymbol{R}^{L-1}}{\partial \boldsymbol{x}^{L-1}} \cdot \boldsymbol{W}^{L-2} \cdot \cdots \cdot \boldsymbol{W}^{l+1} \cdot \frac{\partial \boldsymbol{R}^{l+1}}{\partial \boldsymbol{x}^{l+1}} \cdot \boldsymbol{W}^{l} \cdot \frac{\partial \boldsymbol{R}^{l}}{\partial \boldsymbol{G}^{l}}
\end{aligned}
\end{equation}
where $\boldsymbol{x}^{L}$ is the final output of the RBF-KAN, and $\frac{\partial \mathcal{L}}{\partial \boldsymbol{x}^{L}}$ depends on the choice of the loss function. $\frac{\partial \boldsymbol{R}^{l}}{\partial \boldsymbol{x}^{l}}$ and $\frac{\partial \boldsymbol{R}^{l}}{\partial \boldsymbol{G}^{l}}$ are found by differentiating Equation \ref{RBF_eq}:
\begin{equation}\label{backprop2}
\begin{aligned}
\frac{\partial R^l}{\partial x_i^l} &= -\frac{2(x_i^l - g_j^l)}{\beta^2} \exp \left( - \left( \frac{x_i^l - g_j^l}{\beta} \right)^2 \right)\\
\frac{\partial R^l}{\partial g_j^l} &= \frac{2(x_i^l - g_j^l)}{\beta^2} \exp \left( - \left( \frac{x_i^l - g_j^l}{\beta} \right)^2 \right).\\
\end{aligned}
\end{equation}

\subsection{Neural operators}\label{neural_operator}
\begin{figure}[!htb]
    \centering
    \includegraphics[width=0.6\textwidth]{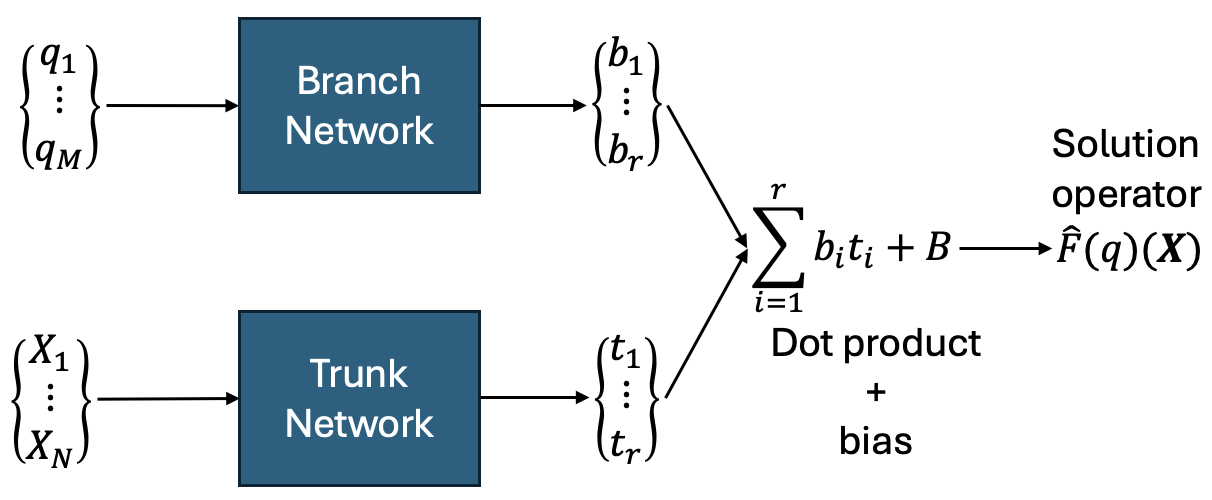}
    \caption{Schematic for neural operators.}
    \label{neural_op_fig}
\end{figure}

Researchers developed neural networks trained to approximate the underlying physics or mathematical operator for a class of problems, a process known as operator learning.  Specifically, neural operators were proposed to learn mapping input functions into corresponding output functions. For neural operators, in the infinite functional space $Q$, the parameter $q \in Q$ represents the input functions and $s \in S$ refers to the unknown solutions of the PDE in the functional space $S$. It is assumed that for each $q$ in $Q$, a unique solution exists $s = s(q)$ in $S$ corresponding to the governing PDE, which also satisfies the boundary conditions (BCs). Consequently, the mapping solution operator $F: Q \rightarrow S$ can be defined as:
\begin{equation}\label{Onet_mapping}
\begin{aligned}
F(q) = s(q).
\end{aligned}
\end{equation}
Specifically, for a collection of $N$ points $\boldsymbol{X}$ on a domain, each denoted by its coordinates $(x_i, y_i)$, the neural operator considers both the functions $q_j$ in its branch and positions $\boldsymbol{X}$ in its trunk. The branch and trunk networks yield outputs with length $r$. The solution operator $\hat{F}(q)(\boldsymbol{X})$ is predicted by fusing intermediate encoded outputs $b_i$ (from branch) and $t_i$ (from trunk) in a dot product enhanced by bias $B$, as shown in Fig. \ref{neural_op_fig}. In a larger sense, it is possible to think of $\hat{F}(q)(\boldsymbol{X})$ as a function of $\boldsymbol{X}$ conditioning on input $q$.

The parameter $q$ can encompass a variety of functional inputs that influence the underlying PDE solution. For instance, $q$ may represent material properties such as thermal conductivity or elastic modulus. Additionally, $q$ could denote load constants or time-dependent input load function, discretized at $T$ time steps to form an input load vector $\boldsymbol{q}$. This allows the modeling of transient processes where the applied loads vary over time, as shown in one of the examples later. This study leverages neural operators as regressors. Thus, one needs to devise an appropriate loss function. Typical loss functions for regression problems are mean square error and mean absolute error. This paper uses the root mean square deviation (RMSD) as the loss function to be minimized and determine the optimized network parameter. The RMSD is defined as: 
\begin{equation}\label{rmsd}
\begin{aligned}
\mathcal{L} = \text{RMSD} = \sqrt{\frac{1}{N} \sum_{i=1}^{N} (s_i - \hat{s}_i)^2}
\end{aligned}
\end{equation}
where $\hat{s}$ denotes the solution obtained from the neural operator. This paper considers two types of neural operators: DeepONet and DeepOKAN.

\subsection{DeepOKAN vs. DeepONet}\label{KAN_deeponet}
\begin{figure}[!htb]
    \centering
    \includegraphics[width=1\textwidth]{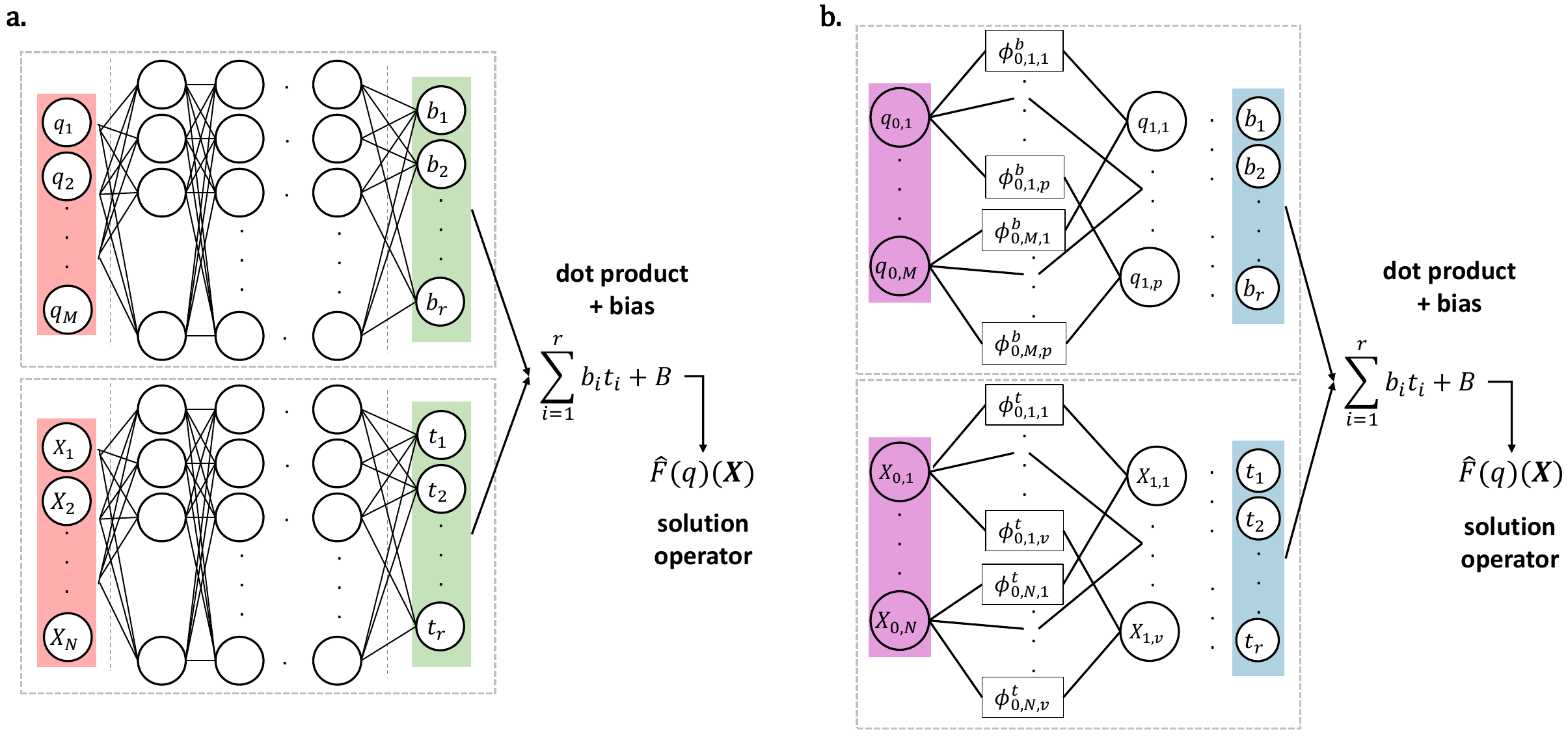}
    \caption{Illustration of ({\bf{a.}}) DeepONets and ({\bf{b.}}) DeepOKANs. $\phi^t$ and $\phi^b$ represent the activation functions corresponding to the trunk and branch networks, respectively. $p$ and $v$ are hyperparameters defining the width of layer $l=1$ in the branch and trunk, respectively. Different layers can possess different widths..}
    \label{Figure_DeepOperators}
\end{figure}

A DeepONet (see Fig. \ref{Figure_DeepOperators}a) model, which uses MLP networks in both its branch and trunk, serves as a performance baseline. This paper introduces the DeepOKAN (see Fig. \ref{Figure_DeepOperators}b), which replaces the MLP networks with KANs. Specifically, it employs Gaussian RBF KANs, though other KAN types, such as multi-quadratic RBF KANs and B-splines KANs, could also be explored in the context of neural operators.

%% file: sections/3Results.tex
\section{Numerical examples}\label{results}

To create a fair evaluation platform for the two operators, DeepOKAN and DeepONet, we adopt the following approach across all the numerical examples. We constrain the depth $d$ of the trunk and branch networks for both operators, as well as the number of neurons in their output layer $r$, and we then vary the number of neurons $n$ in the hidden layers to arrive at the same number of learnable parameters $w$. This approach ensures that essential aspects of the network expressivity, such as the depth, neurons of the output layer, and learnable weights, remain the same across the two architectures. We also maintain identical training settings regarding the choice of batch size, optimizer, number of epochs, and starting learning rate $lr$. The specific implementation details of each example are provided in the following sections. In this study, we use either the 1- L-BFGS optimizer or 2- the Adam optimizer with a learning rate scheduler. The learning rate scheduler is designed to reduce the starting learning rate by a factor of $\gamma$ at regular epochs $T_{step}$, allowing for more precise adjustments as training progresses. This approach helps prevent the model from overshooting the optimal solution. We finally note that throughout the rest of the manuscript and for notation convenience, we will also refer to the number of learnable parameters as \textit{network complexity}, which we denote as $nc$.

\subsection{1D sinusoidal waves}

Before we evaluate the performance of the two operators, DeepOKAN and DeepONet, we compare their baseline networks (RBF-KAN and MLP, respectively). This is an intuitive but important check since we first need to elucidate the behavior of the networks that serve as the backbone of their operator versions. For this task, we use two 1D sinusoidal waves. The equation of the first one is given below: 

\begin{equation}
    y_{1}(x) = \sin(2\pi x) + \cos(\pi x^{2}) + \cos(\pi x^{3}) \times \sin(\pi x^{3})
\label{Wave1}
\end{equation}

\begin{figure}[!htb]
    \centering
    \includegraphics[width=0.8\textwidth]{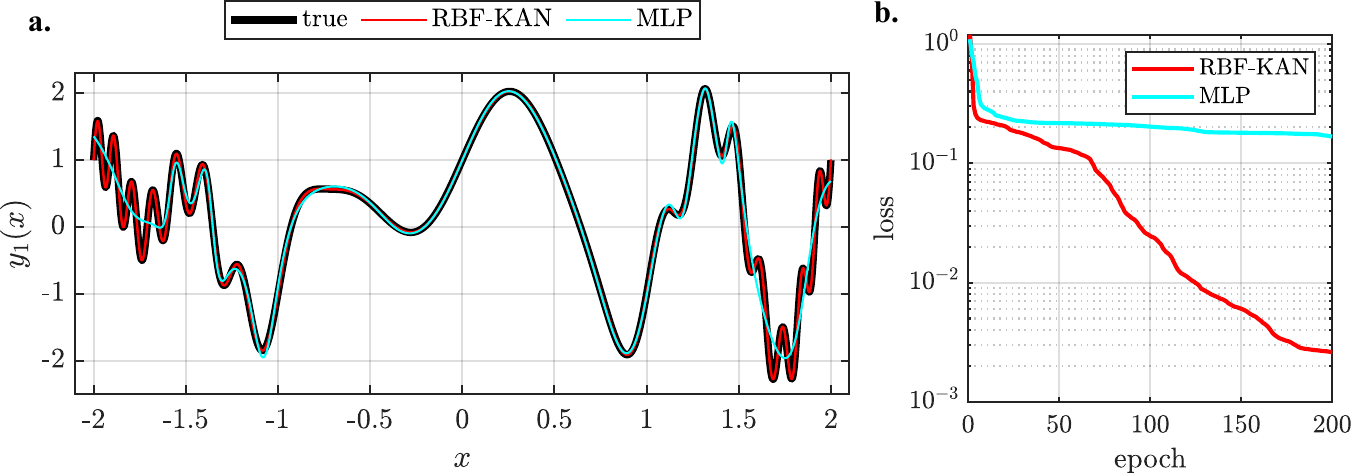}
    \caption{Results of the $1^{st}$ sinusoidal wave example (Wave-Case1): {\bf{a.}} true and predicted values, {\bf{b.}} evolution of the training loss.}
    \label{Figure_Sinewave_Case1}
\end{figure}

Equation \ref{Wave1} is sampled with 1000 uniformly distributed points between [-2, 2]. The RBF-KAN has 2 hidden layers with 8 neurons each, $r = 1$, and $w = 640$ trainable parameters. The MLP is constructed with 2 hidden layers and 24 neurons each, $r = 1$ and $w = 673$ learnable parameters. In this example, we train both networks with the L-BFGS optimizer for 200 epochs, using a learning rate $lr$ = 1 (default) and applying the \textit{tanh()} activation function for the MLP. The results of this study (Wave-Case1) are shown in Fig. \ref{Figure_Sinewave_Case1}, where the left graph compares the true and predicted values from the two networks. We observe that RBF-KAN approximates the true field with much higher accuracy than the MLP, particularly in the high-frequency regions. The improved expressivity of the RBF-KAN is also confirmed by the greater reduction of its loss function compared to the MLP, as shown in Fig. \ref{Figure_Sinewave_Case1}b.

Next, we create a more challenging situation where the wave experiences more intense oscillations. The second wave expression is given below:

\begin{equation}
    y_{2}(x) = \cos(4\pi x) - \sin(\pi x^{2}) \times \cos(\pi x^{3})
\label{Wave2}
\end{equation}

\begin{figure}[!htb]
    \centering
    \includegraphics[width=0.8\textwidth]{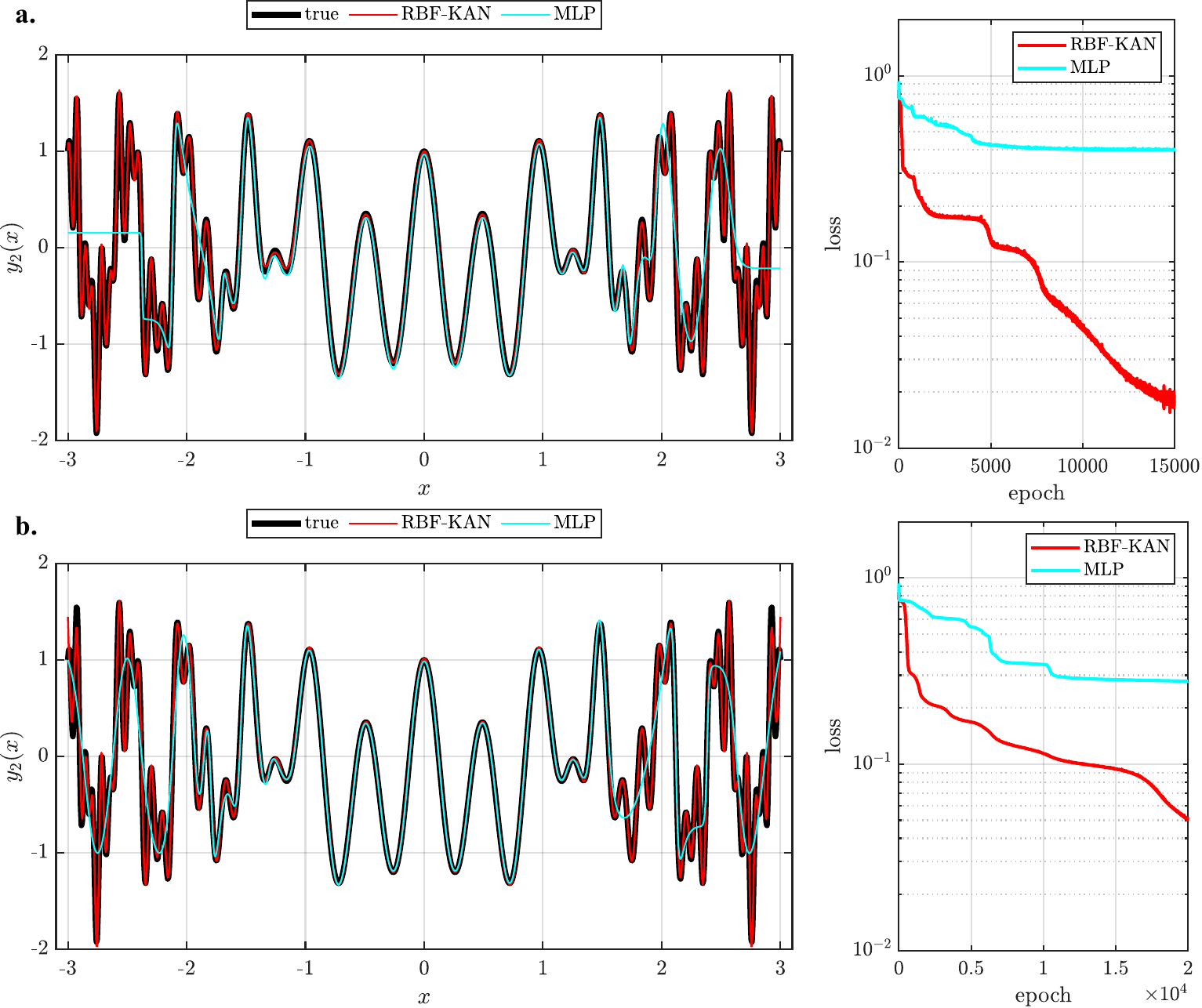}
    \caption{Results of the $2^{nd}$ sinusoidal wave example, training with: {\bf{a.}} Adam, 15000 epochs, $lr = 10^{-2}$ (Wave-Case2A), and {\bf{b.}} Adam, 20000 epochs, $lr = 10^{-3}$ (Wave-Case2B). In both ({\bf{a}}) and ({\bf{b}}), the left graphs show the true and predicted values, and the right graphs depict the evolution of the training loss function.}
    \label{Figure_Sinewave_Case2}
\end{figure}

Equation \ref{Wave2} is sampled with 1000 uniformly distributed points between [-3, 3]. The same architectures for RBF-KAN and MLP are utilized as in the previous case. Here, we explore two training variations. In the first case (Wave-Case2A), we use the Adam optimizer with a learning rate $lr = 10^{-2}$ for 15000 epochs, and in the second case (Wave-Case2B), we use Adam with $lr = 10^{-3}$ for 20000 epochs. The results of this analysis are shown in Fig. \ref{Figure_Sinewave_Case2}. For Wave-Case2A, we observe an almost excellent match between the RBF-KAN and the exact solution, even in the more demanding regions of the spectrum. On the contrary, the MLP fails to capture the high-frequency regimes on both sides of the spectrum. This figure demonstrates the ability of RBF-KAN to approximate the queried field in this problem with much better resolution than the MLP, given that they both have the same number of learnable weights and are trained for the same number of epochs. A similar picture is observed for Wave-Case2A. Here, we note that the match between RBF-KAN and the analytical solution is slightly worse than in the previous case, since there are still some mild discrepancies at the far outermost regimes of the wave. This signifies that training for further epochs could potentially improve the quality of the model predictions, which is also implied by the still descending loss function of RBF-KAN. However, despite the beneficial impact of the lower learning rate on the MLP, it is once again evident that RBF-KAN outperforms MLP in this problem as well. 

Having showcased early signs of the superior performance of RBF-KAN compared to the vanilla MLP architecture, we now proceed with an in-depth comparison of their operator-based versions (DeepOKAN and DeepONet, respectively). The target function is defined as:

\begin{equation}
    y_{2}(x; c) = \cos(c_1 \pi x) - \sin(c_2 \pi x^2) \cos(c_3 \pi x^3),
\label{wave_operator}
\end{equation}

where $x$ ranges from $-3$ to $3$, and the parameters $c_1$, $c_2$, and $c_3$ are sampled from a uniform distribution within the interval $\left[-1, 1\right]$. For the data generation, we created $20,000$ samples. The dataset was subsequently split into training and testing sets, with $80\%$ of the data used for training and the remaining $20\%$ for testing. A batch size of $1024$ is used for both models. The DeepONet and DeepOKAN have 2 layers in the branch and trunk. The DeepOKAN has 50 neurons in each layer, whereas the DeepONet has 350 neurons with $tanh()$ activation. These configurations yield 275880 and 276560 learnable parameters for the DeepONet and DeepOKAN, respectively. Both networks have an $r = 40$ (see Fig. \ref{Figure_DeepOperators}). We consider two independent seeds for the weight initialization of both operators, henceforth termed as $seed1$ and $seed2$. The Adam optimizer with a scheduler is used to optimize the weights of the two models. In this study, we apply the scheduler with $\gamma = 0.9$ every $T_{step}=500$ epochs for a total training epochs of $20,000$. We examine two cases of starting learning rate values, $lr = 10^{-2}$ and $lr = 10^{-3}$.

\begin{figure}[!htb]
    \centering
    \includegraphics[width=0.85\textwidth]{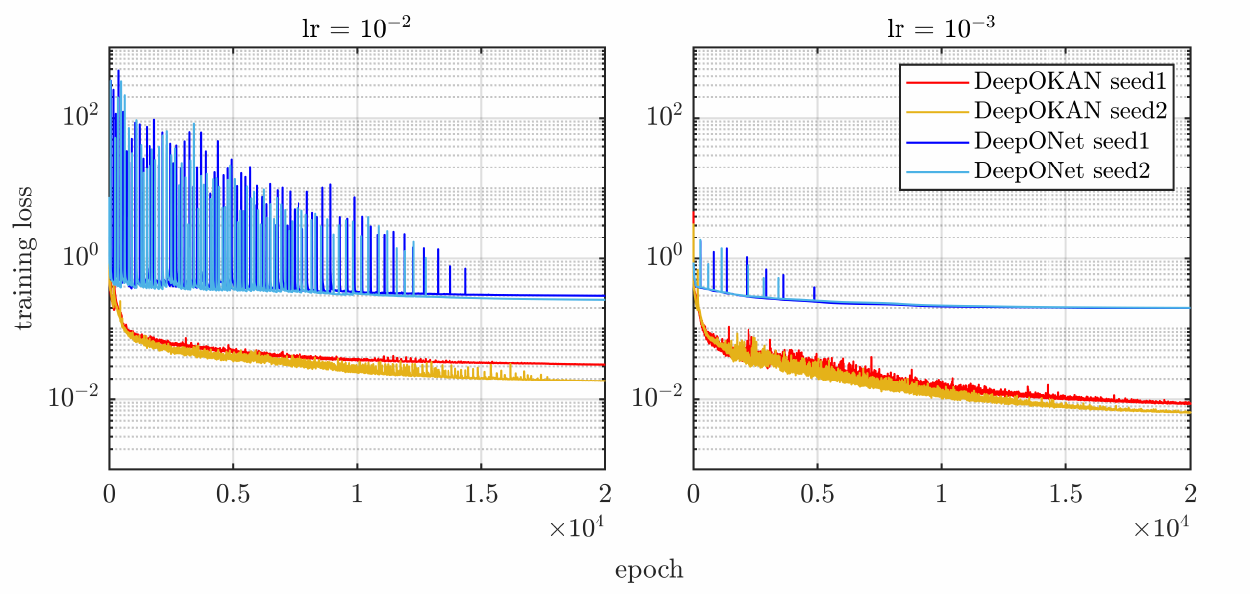}
    \caption{DeepONet and DeepOKAN loss convergence for different seeds and learning rates. Using a starting learning rate of a. $1e-2$ and b. $1e-3$.}
    \label{Figure_1Dwave_Operator_Models_loss}
\end{figure}

The loss convergence histories of all models are shown in Fig. \ref{Figure_1Dwave_Operator_Models_loss}. Overall, the DeepOKAN model demonstrates faster and more stable convergence in both learning rate settings than the DeepONet model. Specifically, for the learning rate of $lr = 10^{-2}$, the DeepONet model exhibits higher spikes and less smooth convergence, indicating potential instability at this higher learning rate. Conversely, the DeepOKAN model consistently converges to a lower training loss faster and with fewer spikes. Both models show improved stability at the lower learning rate of $lr = 10^{-3}$; however, DeepOKAN maintains superior performance. The training loss for DeepOKAN is reduced smoothly and reaches a lower value than the DeepONet models. 

\begin{figure}[!htb]
    \centering
    \includegraphics[width=0.6\textwidth]{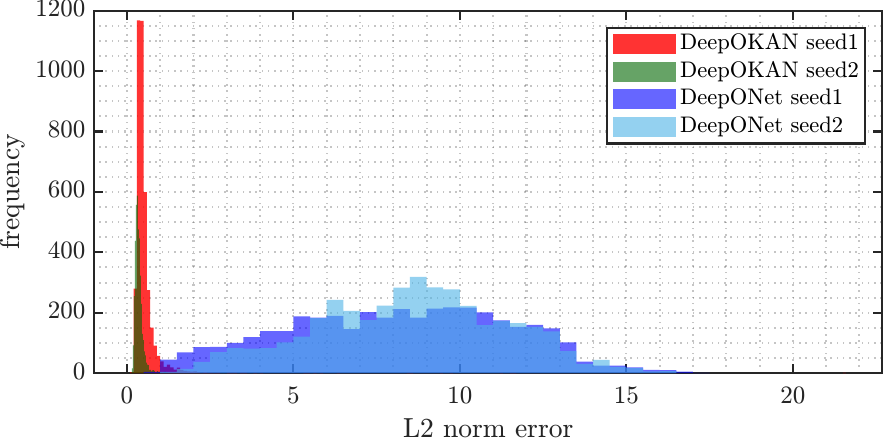}
    \caption{Histogram of the L2-norm values of the absolute errors for the sine wave problem.}
    \label{Figure_1Dwave_Operator_Models_L2graph_histogram}
\end{figure}

\begin{figure}[!htb]
    \centering
    \includegraphics[width=0.95\textwidth]{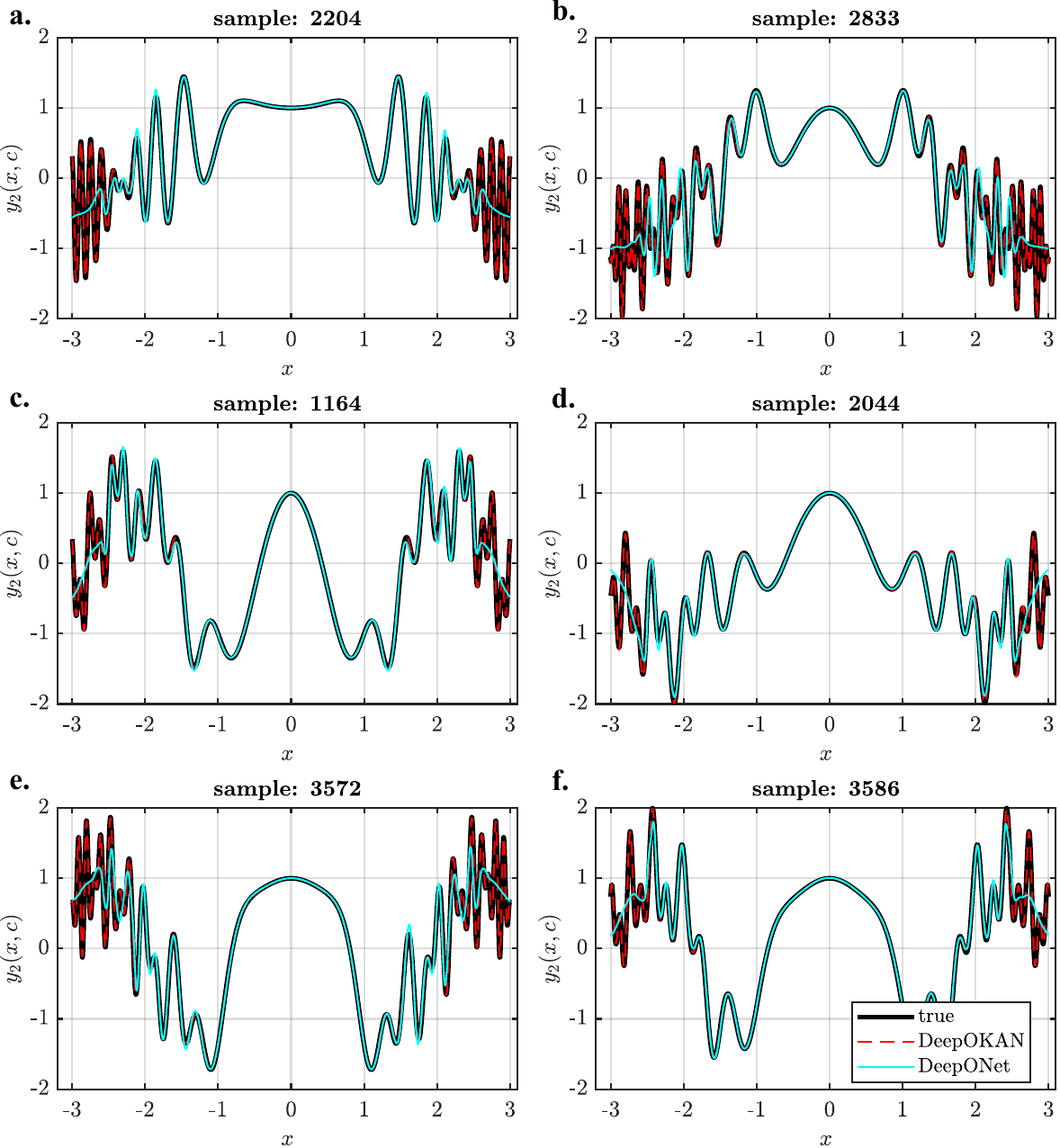}
    \caption{Predictions of DeepONet and DeepOKAN for different samples randomly picked from the testing dataset. The $c$ coefficients for each example are: {(\bf{a.})} c = [0.2276, -0.1843, -0.6270], {(\bf{b.})} c = [0.3257, 0.679, 0.829], {(\bf{c.})} c = [-0.8976, 0.7085, 0.4797], {(\bf{d.})} c = [-0.4736, 0.5641, -0.4129], {(\bf{e.})} c = [-0.7583, -0.5544, -0.7268], {(\bf{f.})} c = [-0.8268, -0.5951, 0.4899].}
    \label{neur_samples}
\end{figure}

The L2-norm of the absolute error values for all the test samples are computed, and they are plotted as shown in the histogram of Fig. \ref{Figure_1Dwave_Operator_Models_L2graph_histogram}. We emphasize here that the same number of samples is plotted in all cases, and overall, this figure illustrates that DeepOKAN's errors are more tightly clustered around smaller values compared to the DeepONet. In particular, the DeepOKAN errors in this case are one order of magnitude smaller than DeepONet. Finally, to further visualize the difference in the performance between the two operators, we plot in Fig. \ref{neur_samples} randomly sampled examples from the testing dataset showing the true functions alongside the predictions from DeepONet and DeepOKAN. Across all samples, DeepOKAN's predictions are consistently closer to the true values, particularly in regions with rapid oscillations and sharp peaks. This indicates its superior ability to capture intricate details and complexities for the 1D sinusoidal wave problem.

\subsection{2D orthotropic elasticity}
\label{ortho_example}

\subsubsection{Problem description, FEA, and data generation}
\label{FEA_ortho}

Next, we consider a homogeneous, orthotropic, elastic body undergoing small deformations. In the absence of body forces and inertial forces, the equilibrium equation can be expressed as follows:
\begin{equation}\label{elasticity_eq}
\begin{aligned}
\boldsymbol{\nabla}\cdot \boldsymbol{\sigma}&=\boldsymbol{0}{,} \quad  \boldsymbol{x}\in\Omega,\\
\boldsymbol{u}&=\overline{\boldsymbol{u}}, \quad \! \boldsymbol{x}\in\Gamma_{u},\\
\boldsymbol{\sigma}\cdot\boldsymbol{n}&=\overline{\boldsymbol{t}}, \quad  \boldsymbol{x}\in\Gamma_{t}.\\
\end{aligned}
\end{equation}
Here, $\boldsymbol{\sigma}$ represents the Cauchy stress tensor, $\boldsymbol{n}$ denotes the normal unit vector, $\boldsymbol{\nabla}\cdot$ signifies the divergence operator, and $\boldsymbol{\nabla}$ indicates the gradient operator. $\Omega$ denotes the domain of the problem, $\Gamma_u$ denotes the part of the boundary where Dirichlet boundary conditions (DBCs) are applied, and $\Gamma_t$ denotes the part of the boundary where Neumann boundary conditions (NBCs) are applied. Given the assumption of small deformations, the strain $\boldsymbol{\varepsilon}$ can be described by:
\begin{equation}\label{eps}
\begin{aligned}
    \boldsymbol{\varepsilon}&=\frac{1}{2}(\boldsymbol{\nabla} \boldsymbol{u}+\boldsymbol{\nabla} \boldsymbol{u}^T)\\
    \end{aligned}
\end{equation}
The relationship between the strain and stress is written as:
\begin{equation}\label{ortho_const}
\begin{aligned}
\boldsymbol{\sigma} &= \mathbb{C} : \boldsymbol{\varepsilon},
\end{aligned}
\end{equation}
where $\mathbb{C}$ is the fourth-order elastic tensor defining linear elastic behavior. In the case of plane-stress orthotropy, it is assumed that the two-dimensional material has two planes of symmetry aligned with the global $x$ and $y$ axes. For the orthotropic plane-stress case, the compliance matrix $\mathbf{S}$ is expressed as: 
\begin{equation}\label{ortho_compliance}
\begin{aligned}
\left[ \mathbf{S} \right] = 
\begin{bmatrix}
\frac{1}{E_x} & -\frac{\nu_{xy}}{E_x} & 0 \\
-\frac{\nu_{yx}}{E_y} & \frac{1}{E_y} & 0 \\
0 & 0 & \frac{1}{G_{xy}}
\end{bmatrix},
\end{aligned}
\end{equation}
with $E_x$ and $E_y$ representing Young's moduli in the orthotropic directions, $G_{xy}$ being the shear modulus, and $\nu_{xy}$ being the in-plane Poisson's ratio. The constitutive relation symmetry is ensured by:
\begin{equation}\label{n_xy}
\begin{aligned}
\nu_{xy}&= \frac{E_{x}}{E_{y}} \nu_{yx}.
\end{aligned}
\end{equation}
$\mathbf{S}$ has to be inverted to obtain the stiffness matrix.

\begin{figure}[!htb]
    \centering
    \includegraphics[width=0.4\textwidth]{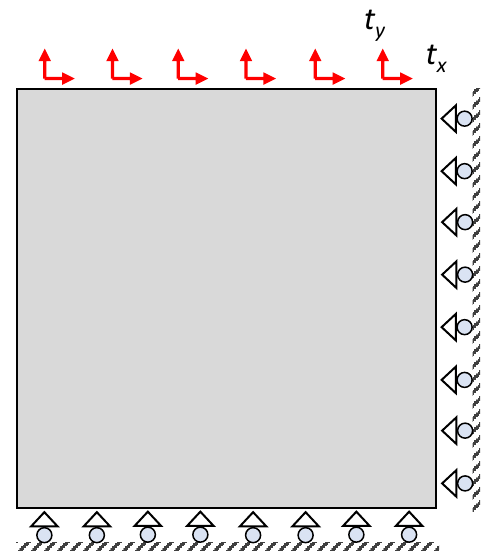}
    \caption{Schematic of the geometry and boundary conditions of the 2D orthotropic material example.}
    \label{Figure_Schematic_Ortho}
\end{figure}

In this example, we consider a two-dimensional (2D) domain subjected to boundary conditions and traction forces shown in Fig. \ref{Figure_Schematic_Ortho}. The bottom and right edges of the domain are subjected to roller boundary conditions. Traction forces are applied at the top edge of the domain, represented by two components, $t_x$ and $t_y$. These forces are sampled from uniform distributions within a specified range: $t_x$, $t_y$ $\in$ $\left[-0.3,0.3\right]$. As demonstrated in Equations \ref{ortho_compliance} and \ref{n_xy}, for a 2D orthotropic material, four constants are required: $E_x$, $E_y$, $\nu_{xy}$, and $G_{xy}$. We sample $E_x$, $E_y$, and $\nu_{xy}$ from specified uniform distributions: $E_{x}, E_{y} \in \left[5,20\right]$, and $\nu_{xy} \in \left[0.15, 0.35\right]$. $G_{xy}$ is computed using the following relationship:
\begin{equation}\label{G_xy_generated}
\begin{aligned}
G_{xy}&= \frac{1}{2}\left(\frac{E_x}{2\left(1+\nu_{xy}\right)} + \frac{E_y}{2\left(1+\nu_{yx}\right)}\right).
\end{aligned}
\end{equation}

Equation \ref{G_xy_generated} does not generally hold for orthotropic materials. However, we compute $G_{xy}$ using this equation to ensure that it remains within a physically plausible range, as for most materials, it is observed that they are less stiff in shear than in tension or compression. Hence, we would like to highlight that it is an assumption we made, and one might choose a different range to generate the dataset for training. Although $G_{xy}$ is calculated using Equation \ref{G_xy_generated} in the data generation phase, it is still fed to the neural operators to learn its impact on the material deformation. We assume zero body force and negligible inertia. This leaves us with six parameters to be fed to the branch network: $t_x$, $t_y$, $E_x$, $E_y$, $\nu_{xy}$, and $G_{xy}$, while the coordinates are fed to the trunk network. A total of 5000 samples were generated using this approach. The dataset was then split into 80\% for training and 20\% for testing.

\subsubsection{Training and Results}\label{results_ortho}

In this example, we investigate three cases of network complexity: $nc = [low, \ medium, \ high]$. Both DeepOKAN and DeepONet have a single hidden layer in their trunk and branch networks, as well as 5 neurons in the output layer. Across the three levels of complexity, the DeepOKAN has $n = [14, \ 80, \ 190]$ neurons in its hidden layer and $w = [1260, \ 7200, \ 17100]$ trainable parameters. Respectively, the DeepONet has $n = [62, \ 358, \ 855]$ neurons and $w = [1250, \ 7170, \ 17110]$ parameters. Both operators are trained with Adam for 10000 epochs, using a learning rate $lr = 10^{-3}$ and a batch size of 64. A learning scheduler is applied with $\gamma = 0.5$ and $T_{step}=1000$ epochs to enhance the training process. Finally, we perform two independent simulations for each operator, using two different seed numbers to initialize their weights ($seed1$ and $seed2$). 

\begin{figure}[!htb]
    \centering
    \includegraphics[width=1\textwidth]{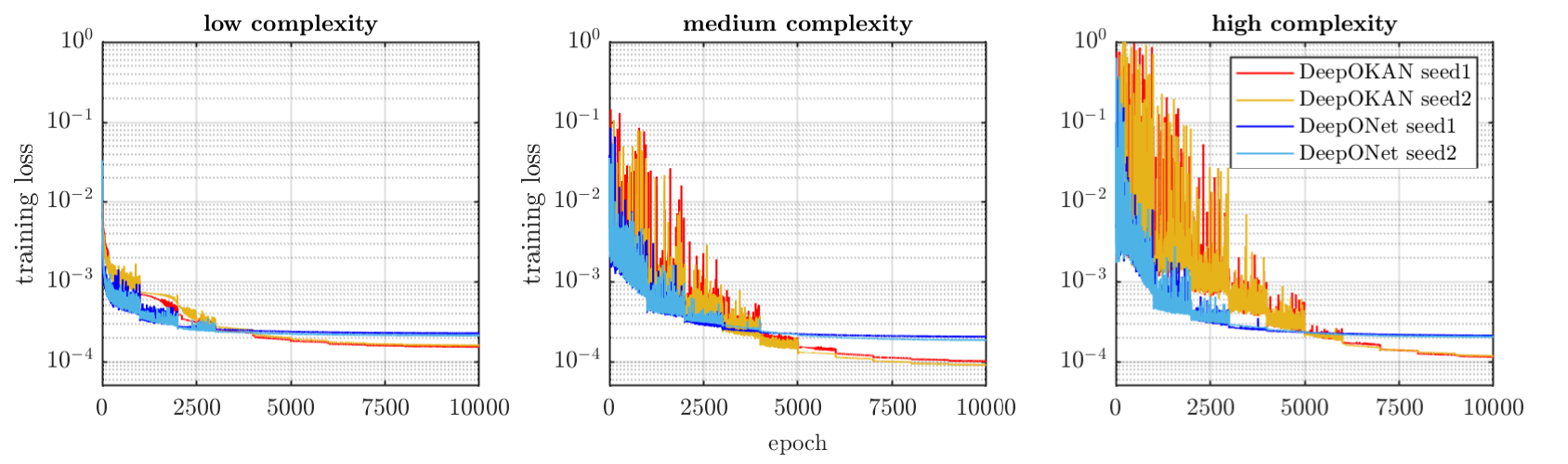}
    \caption{Evolution of the training loss function for the 2D orthotropic elasticity problem.}
    \label{Figure_Ortho_Models_loss}
\end{figure}

\begin{figure}[!htb]
    \centering
    \includegraphics[width=1\textwidth]{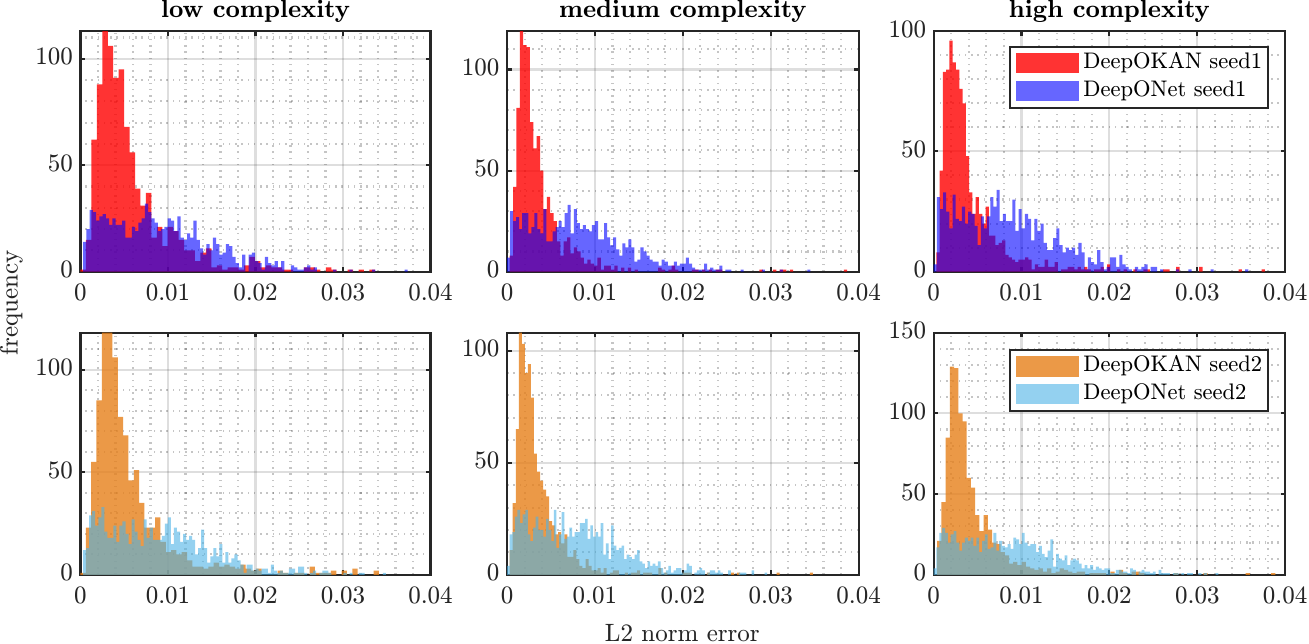}
    \caption{Histograms of the L2-norm values of the absolute errors for the 2D orthotropic elasticity problem.}
    \label{Figure_Ortho_Models_L2norms}
\end{figure}

The evolution of the training loss functions for these models is shown in Fig. \ref{Figure_Ortho_Models_loss}. Across all levels of complexity, we observe that DeepONets experience a sharper decrease in their loss function at the beginning of their training. However, the learning capacity of the network seems to be stalled rather quickly, as evidenced by the almost negligible loss reduction after a few thousand epochs. On the contrary, all DeepOKAN models experience a smoother reduction in the training loss function, ultimately arriving at smaller values at the end of the training. This is the first indication that DeepOKAN may also bear higher accuracy than its DeepONet counterparts for this problem. We compute the L2-norm of the absolute error values for all the test samples and plot their histograms in Fig. \ref{Figure_Ortho_Models_L2norms} to evaluate this assumption. For visual clarity, the top row of graphs represents the $seed1$ models, while the bottom graphs represent $seed2$. Here, we observe that DeepOKANs attain smaller values in the L2-norm, clearly outperforming their DeepONets counterparts. This observation holds true for both seed cases, regardless of the level of complexity.

\begin{figure}[!htb]
    \centering
    \includegraphics[width=1\textwidth]{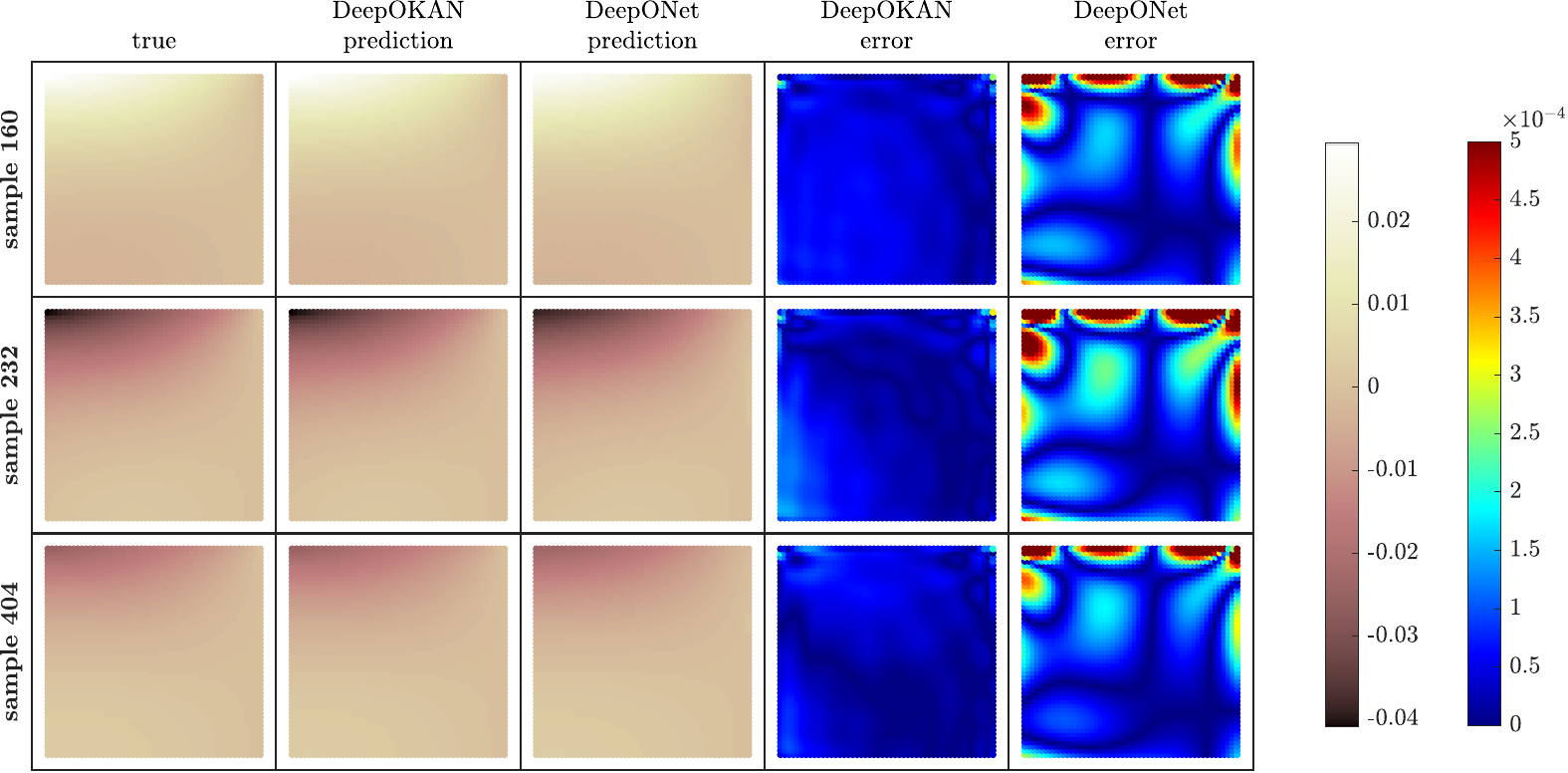}
    \caption{Comparison between true and predicted values for the 2D orthotropic elasticity problem.}
    \label{Figure_Ortho_Maps_M2_seed1}
\end{figure}

To further solidify these observations, we randomly select three test samples and in Fig. \ref{Figure_Ortho_Maps_M2_seed1} we plot their true values, predictions and absolute error maps. The medium complexity DeepOKAN and DeepONet with $seed1$ are chosen for evaluation purposes. Even though both operators capture the true field with sufficient accuracy, both qualitatively and quantitatively, DeepOKANs are clearly more accurate than DeepONets, as shown by the absolute error contour maps. Here we emphasize that the colorbar of the absolute error graphs is capped between 0 and 0.0005. 


\subsection{Transient Poisson's problem}
\label{poisson_example}

\subsubsection{Problem description, FEA, and data generation}
\label{poisson_fea}

\begin{figure}[!htb]
    \centering
    \includegraphics[width=1\textwidth]{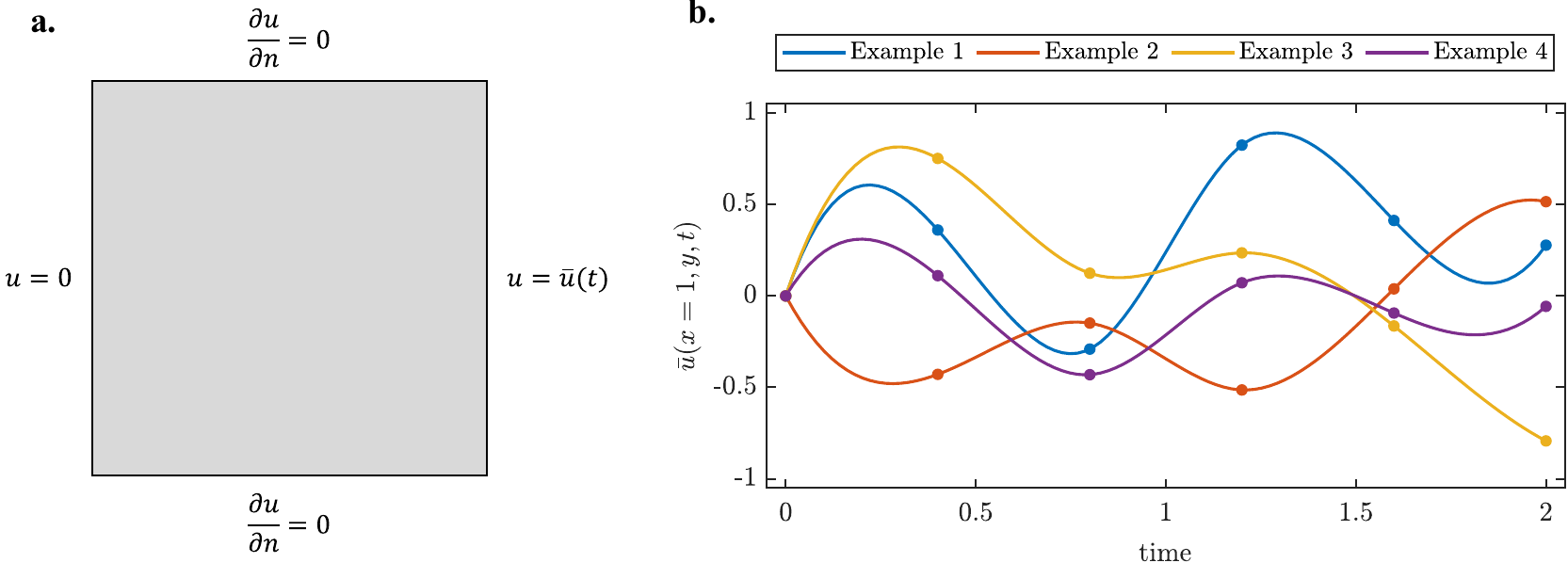}
    \caption{{\bf{a.}} Schematic of the geometry and boundary conditions of the 2D transient Poisson example. {\bf{b.}} Examples of the DBCs imposed on the right edge of the unit square domain. The points appearing on the curves represent the control points sampled from the normal distribution and then used for interpolation.}
    \label{Figure_Transient_Schematic}
\end{figure}

In this example, we will demonstrate how the DeepOKAN can solve the transient Poisson problem. Consider a homogeneous unit square experiencing a time-dependent DBC at the right edge, while zero DBC is imposed at the left edge of the square. The top and bottom edges experience zero NBCs. Fig. \ref{Figure_Transient_Schematic}a summarizes the domain and boundary conditions used for this example. This transient Poisson problem can be described by the partial differential equation (PDE):
\begin{equation}\label{poisson}
\begin{aligned}
\frac{\partial u}{\partial t} - \Delta u &= f  \quad \; \; \; \; \! \quad \boldsymbol{x}\in\Omega \text{, } t\in (0, T],\\
u(x, 0) &= u_0(x) \quad  \; \boldsymbol{x}\in\Omega\\
u &= \overline{u}\left(t\right) \; \; \; \quad \! \boldsymbol{x}\in\Gamma_{u}\\
\nabla u \cdot \mathbf{n} &= 0 \quad \; \; \; \quad \; \boldsymbol{x} \in \Gamma_t
\end{aligned}
\end{equation}
where $\Delta$ is the Laplacian, and $f$ is the source term, set to zero for this example. $u_0(x)$ is the initial condition, and it is set to zero, i.e., $u_0(x)=0$.

To generate the dataset for training and testing neural operators on the transient Poisson problem with a time-dependent Dirichlet boundary condition applied to the right edge of a unit square domain, we begin by randomly sampling five control points. These control points are generated by first creating a sequence of time stamps linearly spaced between the initial time $t=0$ and the final time $t=T$ and then sampling corresponding $u$ values from a normal distribution $\mathcal{N}(0, \sigma = 0.5)$, which are clipped to remain within the range $\left[-1,1\right]$. The initial condition at $t=0$ with $u=0$ is explicitly included. The complete set of six points (origin plus five randomly sampled points) is then interpolated using a cubic spline function, denoted as $CS\left(t\right)$, to ensure a smooth transition between these points. Mathematically, this can be expressed as:
\begin{equation}\label{cubic_spline_poisson}
\begin{aligned}
CS(t) &= \sum_{i=0}^{n} c_i B_i(t)
\end{aligned}
\end{equation}
where $B_{i}(t)$ are the basis functions of the cubic spline, and $c_{i}$ are the coefficients determined by the control points. This interpolation is evaluated at $100$ evenly spaced time points between $t=0$ and $t=T$ to generate a time series of $u$ values. Fig. \ref{Figure_Transient_Schematic}b shows a few examples of the generated boundary conditions imposed on the right edge of the domain.

Next, these time-dependent DBCs are used in the FEA to solve for the corresponding transient solutions. The FEA is performed using FEniCSx. A total of 4500 samples were generated using this approach. The dataset was then divided into 80\% for training and 20\% for testing. The branch network takes the boundary condition history at the right edge $\overline{u}\left(x=1, y, t\right)$. The length of the branch input vector is $M=100$, where each entry corresponds to one point in time. On the other hand, the trunk takes the coordinates of the domain. Since we are interested in transient response at the $M$ time increments, we must devise the branch output to reflect such a dimensionality. The branch output has a length of $M\times r$. Then, we reshape the output to $\left(M, r\right)$ to prepare it for the multiplication with the trunk output with a length of $r$. Unlike the previous examples, the multiplication yields an output of length $M$ representing the solution history: 
\begin{equation}\label{trans_operator}
\begin{aligned}
\hat{F}(q)(\boldsymbol{X})=\sum_{i=1}^{r} b_{m i} t_{i} + B_{m}\\
\end{aligned}
\end{equation}
where $m=[1, 2, \dots, M]$ is the time increment.

\subsubsection{Training and Results}\label{poisson_results}

\begin{figure}[!htb]
    \centering
    \includegraphics[width=1.0\textwidth]{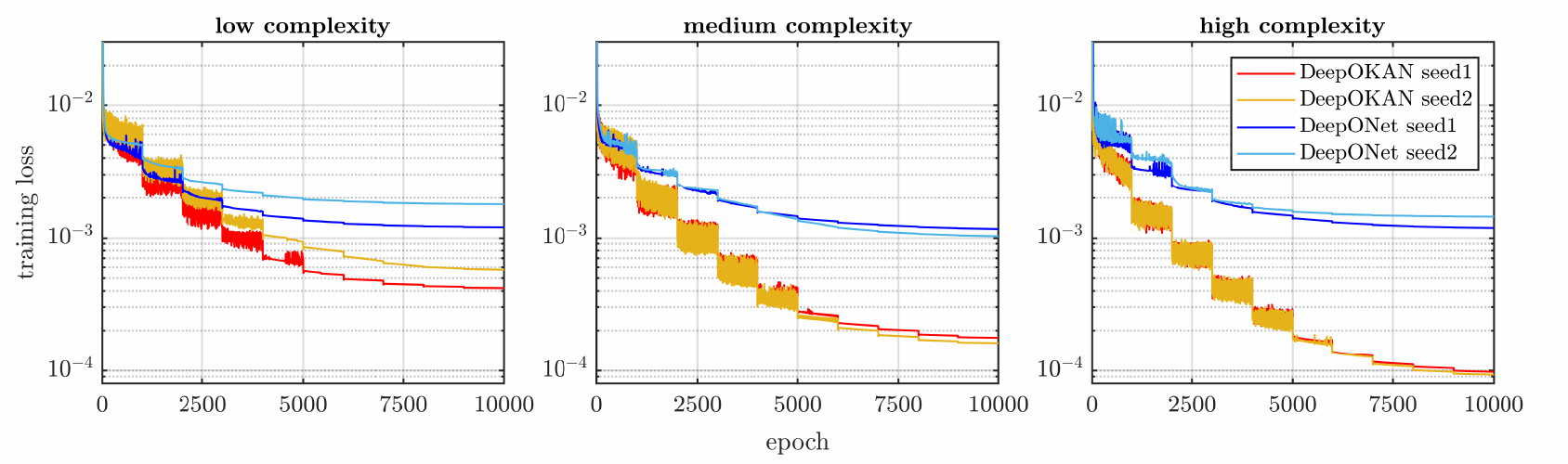}
    \caption{Training loss for the transient Poisson problem: Two different seeds for each network complexity level.}
    \label{Figure_TransientPoisson_loss}
\end{figure}

In this example, we consider three cases of network complexity: $nc=$ low, $nc=$ medium, and $nc=$ high. For the low complexity case, the DeepOKAN has $n=5$ neurons in the hidden layer, while the DeepONet has $n=25$. This leads to $w=13104$	and $w=12650$ learnable parameters for the DeepONet and DeepOKAN, respectively. For the medium complexity case, the DeepOKAN has $n=10$ with $w=25300$ learnable parameters, while the DeepONet has $n=50$ with $w=25804$ learnable parameters. For the high complexity case, $n=20$ are assigned to the DeepOKAN, leading to $w=50600$ learnable parameters, and $n=100$ are assigned to the DeepONet, yielding $w=51204$ learnable parameters. In all cases, $r$ is set to $r=4$. Fig. \ref{Figure_TransientPoisson_loss} presents training loss curves for the DeepOKAN and DeepONet across low, medium, and high complexity levels. Each subplot displays the training loss over epochs for two seeds for each neural operator. DeepOKAN consistently achieves lower training losses than DeepONet. This trend is evident across low, medium, and high complexity levels, indicating DeepOKAN's superior training efficiency and performance stability regardless of complexity.

\begin{figure}[!htb]
    \centering
    \includegraphics[width=0.8\textwidth]{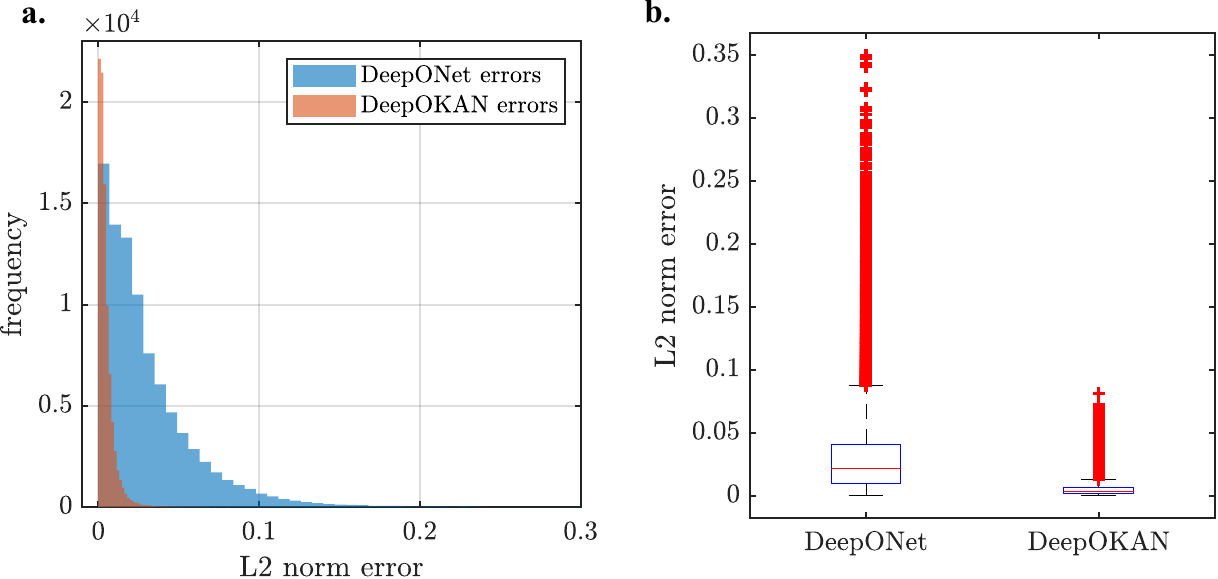}
    \caption{Histogram and box plot of L2-norm errors for DeepONet and DeepOKAN on the testing dataset.}
    \label{Figure_Transient_Poisson_L2norms}
\end{figure}

\begin{figure}[!htb]
    \centering
    \includegraphics[width=0.85\textwidth]{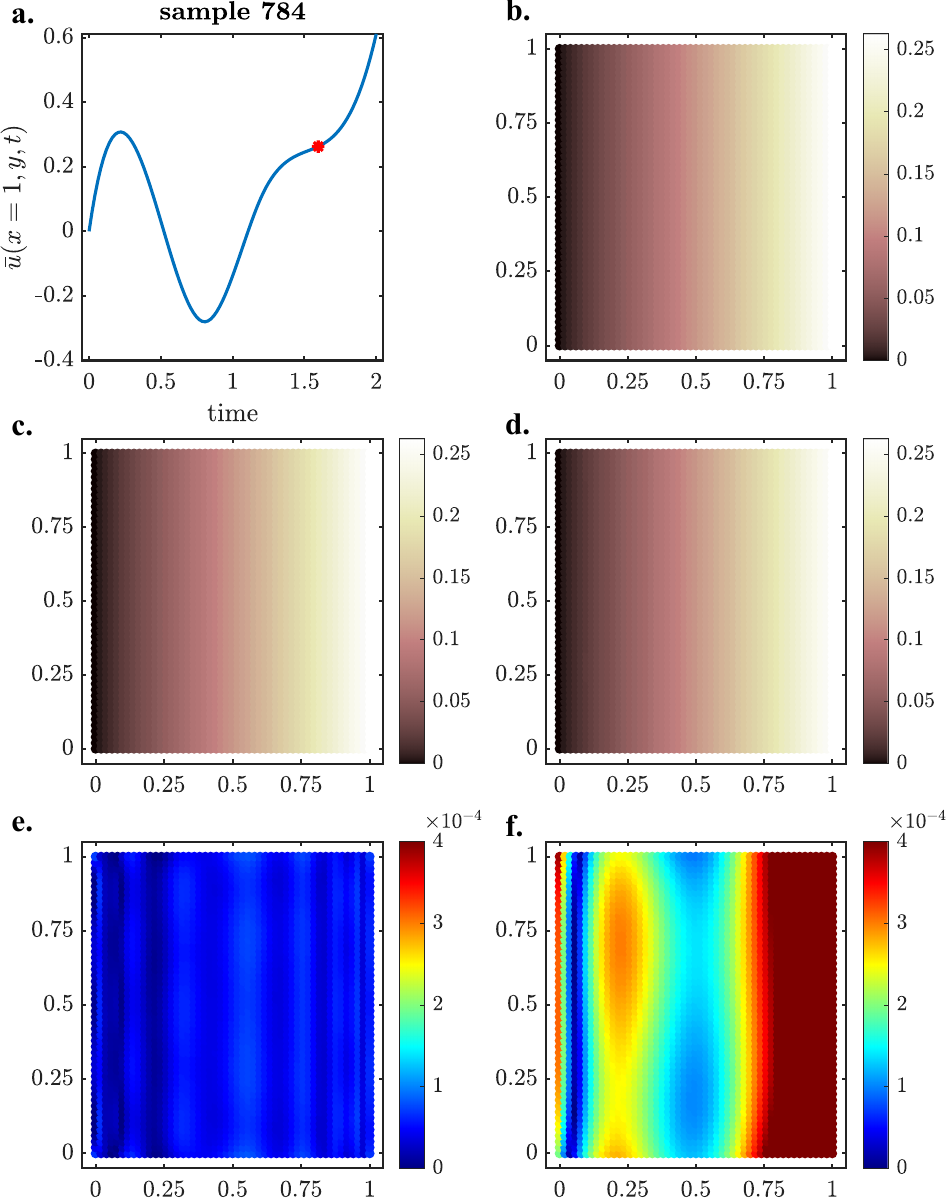}
    \caption{Comparison between the solutions and errors of the DeepONet and DeepOKAN. The contours are for a randomly selected point from the testing dataset and time increment (shown as a red point in a.). a. DBC history imposed on the right edge of the unit square domain. b. Ground-truth solution obtained using the FEA. c. DeepOKAN prediction. d. DeepONet prediction. e. DeepOKAN error. f. DeepONet error.}
    \label{Figure_TransientPoisson_maps_2_numbered}
\end{figure}

\begin{figure}[!htb]
    \centering
    \includegraphics[width=0.85\textwidth]{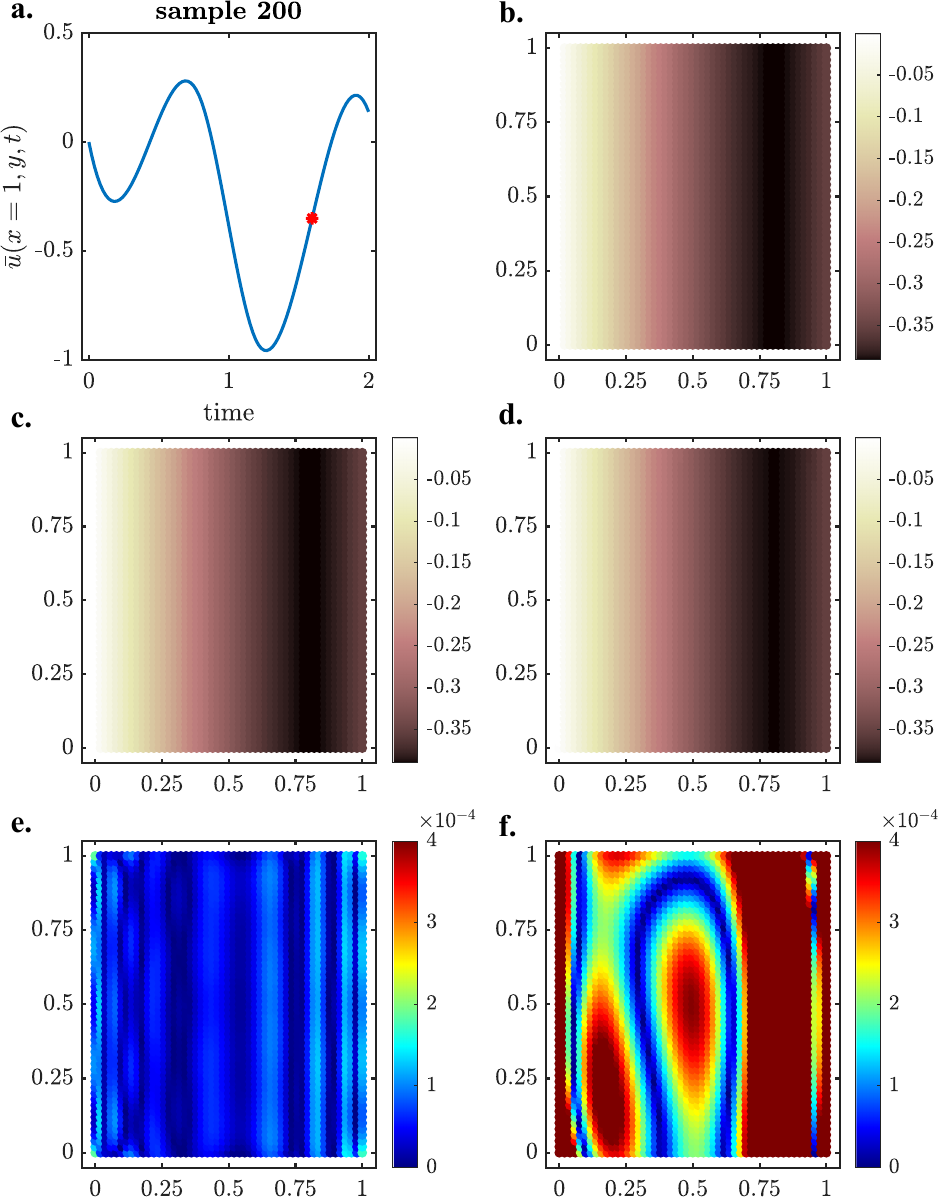}
    \caption{Comparison of DeepOKAN and DeepONet for another test example. The figure layout is identical to Fig. \ref{Figure_TransientPoisson_maps_2_numbered}}
    \label{Figure_TransientPoisson_maps_1_numbered}
\end{figure}

Next, we perform a statistical analysis for the DeepONet and DeepOKAN using the testing dataset, where the analysis is done using the high-complexity network of each model. The analysis reveals that DeepOKAN consistently outperforms DeepONet. This is evident from Fig. \ref{Figure_Transient_Poisson_L2norms}a and Table \ref{errors_stats}. Table \ref{errors_stats} presents the statistics: the mean L2-norm error for DeepONet is $0.02980$ with a standard deviation of $0.0302$, whereas DeepOKAN has a mean error of $0.0047$ and a standard deviation of $0.0052$. The median, $25^{th}$ percentile, and $75^{th}$ percentile values also highlight DeepOKAN's superior performance. Fig. \ref{Figure_Transient_Poisson_L2norms}b illustrates that DeepONet has more outliers and greater variability, indicating less stable performance. The histogram further demonstrates that DeepOKAN's errors are more tightly clustered around smaller values. Finally, Figs. \ref{Figure_TransientPoisson_maps_2_numbered} and \ref{Figure_TransientPoisson_maps_1_numbered} compare the solutions obtained from the DeepONet and DeepOKAN and their corresponding errors for two sample examples. Altogether, these figures further manifest the conclusion that DeepOKAN outperforms the DeepONet.

\begin{table}[!htb]
    \centering
    \caption{Statistical analysis of L2-norm errors for the DeepONet and DeepOKAN.}
    \scalebox{0.9}{ 
        \begin{tabular}{lcccccc}
            \hline
            Network & Mean & Std Deviation & Median & $25^{th}$ Percentile & $75^{th}$ Percentile \\
            \hline
            DeepONet & 0.02980 & 0.0302 & 0.0214 & 0.0099 & 0.0408 \\
            DeepOKAN & 0.0047 & 0.0052 &  0.0033 & 0.0016 & 0.0061 \\
            \hline
        \end{tabular}
    }
    \label{errors_stats}
\end{table}

%% file: sections/4Conclusions.tex
\section{Conclusions, limitations, and future work}\label{conclu}

Neural operators are machine learning models that approximate the solutions of partial differential equations by learning the mapping between function spaces, providing an efficient and accurate tool for solving complex, high-dimensional problems in physics and engineering. One of the most popular neural operators is the DeepONet, which utilizes the Multi-Layer Perceptron (MLP) architecture in the branch and trunk, where the MLP is based on interleaving affine transformations and nonlinear activation functions. MLPs treat the linear transformations and activation functions separately. This paper proposes a new neural operator based on KANs called DeepOKAN. The proposed operator uses KANs in the branch and trunk components, where KANs have learnable activation functions on the edges of the network and sum operations on the nodes. One may consider different bases for these activation functions. This paper uses Gaussian RBFs as a basis for the KANs, appearing in both the branch and trunk. The classical data-driven DeepONet framework is effective; however, the proposed DeepOKAN outperforms it.

%% file: sections/Acknowledgment.tex
\section*{Acknowledgment}
This work was partially supported by the Sand Hazards and Opportunities for Resilience, Energy, and Sustainability (SHORES) Center, funded by Tamkeen under the NYUAD Research Institute Award CG013. The authors wish to thank the NYUAD Center for Research Computing for providing resources, services, and skilled personnel. This work was partially supported by Sandooq Al Watan Applied Research and Development (SWARD), funded by Grant No.: SWARD-F22-018.

%% file: sections/DataAvail.tex
\section*{Data availability}
The data supporting the study's findings will be available upon paper acceptance.